# Sciduction: Combining Induction, Deduction, and Structure for Verification and Synthesis[*]

Sanjit A. Seshia
University of California, Berkeley
sseshia@eecs.berkeley.edu

## Abstract

Even with impressive advances in automated formal methods, certain problems in system verification and synthesis remain challenging. Examples include the verification of quantitative properties of software involving constraints on timing and energy consumption, and the automatic synthesis of systems from specifications. The major challenges include environment modeling, incompleteness in specifications, and the complexity of underlying decision problems.

This position paper proposes sciduction, an approach to tackle these challenges by integrating *inductive inference*, *deductive reasoning*, and *structure hypotheses*. Deductive reasoning, which leads from general rules or concepts to conclusions about specific problem instances, includes techniques such as logical inference and constraint solving. Inductive inference, which generalizes from specific instances to yield a concept, includes algorithmic learning from examples. Structure hypotheses are used to define the class of artifacts, such as invariants or program fragments, generated during verification or synthesis. Sciduction constrains inductive and deductive reasoning using structure hypotheses, and actively combines inductive and deductive reasoning: for instance, deductive techniques generate examples for learning, and inductive reasoning is used to guide the deductive engines.

We illustrate this approach with three applications: (i) timing analysis of software; (ii) synthesis of loop-free programs, and (iii) controller synthesis for hybrid systems. Some future applications are also discussed.

## 1 Introduction

Formal verification has made enormous strides over the last few decades. Verification techniques such as model checking [12, 45, 14] and theorem proving (see, e.g. [26]) are used routinely in computer-aided design of integrated circuits and have been widely applied to find bugs in software. However, certain problems in system verification and synthesis remain very challenging, stymied by computational hardness or requiring significant human input into the verification process. This position paper seeks to outline these challenges and presents an approach for tackling them along with some initial evidence for the utility of the approach.

Let us begin by examining the traditional view of verification is as a decision problem, with three inputs (see Figure 1):

1. A model of the system to be verified, *S*;

2. A model of the environment, *E*, and

3. The property to be verified, Φ.

---

[*]This is a revised version of a previously-published technical report [51].

The verifier generates as output a YES/NO answer, indicating whether or not S satisfies the property Φ in environment E. Typically, a NO output is accompanied by a counterexample, also called an error trace, which is an execution of the system that indicates how Φ is violated. Some formal verification tools also include a proof or certificate of correctness with a YES answer.

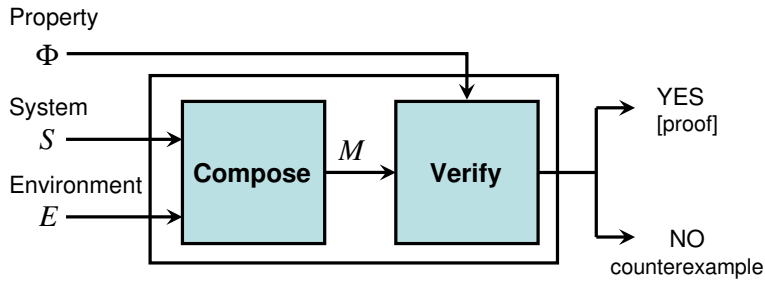

Figure 1: Formal verification procedure.

The first point to note is that this view of verification is somewhat high-level and idealized. In practice, one does not always start with models S and E — these might have to be generated from implementations. To create the system model S, one might need to perform automatic abstraction from code that has many low-level details. For this, techniques such as predicate abstraction [21] have proved useful (e.g., in software model checking [4]). Additionally, the specification Φ is rarely complete and sometimes inconsistent, as has been noted in industrial practice (see, e.g., [6]). Indeed, the question "when are we done verifying?" often boils down to "have we written enough properties (and the right ones)?" Finally, the generation of an environment model E is usually a manual process, involving writing constraints over inputs, or an abstract state machine description of the environment of S. Bugs can be missed due to incorrect environment modeling. In systems involving commercial-off-the-shelf components or third-party IP blocks, not all details of the environment might even be available. As an example of the challenge of environment modeling, consider the verification of quantitative properties, such as bounds on timing and power. Such analysis is central to the design of reliable embedded software and systems, such as those in automotive, avionics, and medical applications. However, the verification of such properties on a program is made difficult by their heavy dependence on the program's environment, which includes (at a minimum) the processor it runs on and characteristics of the memory hierarchy. Current practice requires significant manual modeling, which can be tedious, error-prone and time consuming, taking several months to create a model of a relatively simple microcontroller.

The second point we note is that Figure 1 omits some inputs that are key in successfully completing verification. For example, one might need to supply hints to the verifier in the form of inductive invariants or pick an abstract domain for generating suitable abstractions. One might need to break up the overall design into components and construct a compositional proof of correctness (or show that there is a bug). These tasks requiring human input have one aspect in common, which is that they involve a *synthesis sub-task* of the overall verification task. This sub-task involves the synthesis of *verification artifacts* such as inductive invariants, abstractions, environment assumptions, input constraints, auxiliary lemmas, etc. One often needs human insight into at least the form of these artifacts, if not the artifacts themselves, to succeed in verifying the design.

Finally, it has been a long-standing goal of the fields of electrical engineering and computer science to automatically synthesize systems from high-level specifications. Automatic synthesis shares much in common with automatic verification. It is interesting to note that the genesis of model checking lies in part in the automatic synthesis problem; the seminal paper on model checking by Clarke and Emerson [12] begins with this sentence:

"*We propose a method of constructing concurrent programs in which the synchronization skeleton*

*of the program is automatically synthesized from a high-level (branching time) Temporal Logic specification."*

In automatic synthesis, one starts with a specification $\Phi$ of the system to be synthesized, along with a model of its environment $E$. The goal of synthesis is to generate a system $S$ that satisfies $\Phi$ when composed with $E$. Figure 2 depicts the synthesis process. Modeled thus, the essence of synthesis can be viewed as a *game*

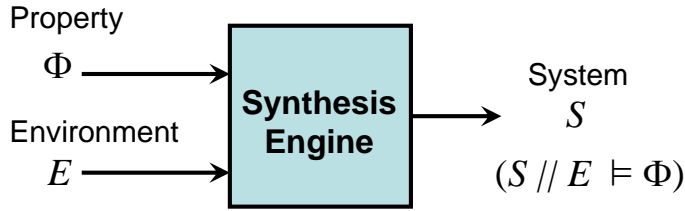

Figure 2: Formal synthesis procedure.

*solving* problem, where $S$ and $E$ represent the two players in a game; $S$ is computed as a winning strategy ensuring that the composed system $S\|E$ satisfies $\Phi$ for all input sequences generated by the environment $E$. If such an $S$ exists, we say that the specification ($\Phi$ and $E$) is *realizable*. Starting with the seminal work on automata-theoretic and deductive synthesis from specifications (e.g. [33, 42]), there has been steady progress on automatic synthesis. In particular, many recent techniques (e.g. [57, 59]) leverage the progress in formal verification in order to perform synthesis. However, there is a long way to go before automated synthesis is practical and widely applicable. One major challenge, shared with verification, is the difficulty of obtaining complete, formal specifications from the user. Even expert users find it difficult to write complete, formal specifications that are realizable. Often, when they do write complete specifications, the effort to write these is arguably more than that required to manually create the design in the first place. Additionally, the challenge of modeling the environment, as discussed above for verification, also remains for synthesis. Finally, synthesis problems typically have greater computational complexity than verification problems for the same class of specifications and models. For instance, equivalence checking of combinational circuits is NP-complete and routinely solved in industrial practice, whereas synthesizing a combinational circuit from a finite set of components is $\Sigma_2$-complete and only possible in very limited settings in practice. In some cases, both verification and synthesis are undecidable, but there are still compelling reasons to have efficient procedures in practice; a good example is hybrid systems — systems with both discrete and continuous state — whose continuous dynamics is non-linear, which arise commonly in embedded systems and analog/mixed-signal circuits.

To summarize, there are two main points. First, the core challenges facing automatic verification are very similar to those facing automatic synthesis: the key to efficient verification is often in the synthesis of artifacts such as inductive invariants or abstractions, and synthesizers employ verifiers, at a minimum to verify that the systems they generate are correct. Second, the three main challenges facing formal verification and synthesis are: (i) system and environment modeling, (ii) creating good specifications, and (iii) the complexity of underlying decision problems. Some of these challenges — such as dealing with computational complexity — can be partially addressed by advances in *computational engines* such as Binary Decision Diagrams (BDDs) [9], Boolean satisfiability (SAT) [32], and satisfiability modulo theories (SMT) solvers [5]. However, these alone are not sufficient to extend the reach of formal methods for verification and synthesis. New *methodologies* are also required.

In this position paper, we present one such methodology that, in our experience so far, appears promising. The central idea of this approach, which we term *sciduction*,[1] is to tightly integrate *induction*, *deduction*, and

---
[1] sciduction stands for structure-constrained induction and deduction.

*structure hypotheses*. *Induction* is the process of inferring a general law or principle from observation of particular instances. Machine learning algorithms are typically inductive, generalizing from (labeled) examples to obtain the learned concept or classifier [38, 2]. *Deduction*, on the other hand, involves the use of rules and axioms to logically infer conclusions about particular problem instances. Traditional approaches to formal verification and synthesis, including those in model checking and theorem proving, tend to be deductive. In formal verification and synthesis (see Figures 1 and 2), given a particular specification $\Phi$, environment $E$, and system $S$, a verifier typically uses a combination of search and logical inference designed for that class of $\Phi$, $E$, and $S$ to determine if $S\|E \models \Phi$. On the other hand, inductive reasoning may seem out of place here, since typically an inductive argument only ensures that the truth of its premises make it *likely or probable* that its conclusion is also true. However, we argue in this paper that one can combine inductive and deductive reasoning to obtain the kinds of guarantees one desires in formal verification and synthesis. The key is the use of *structure hypotheses*. These are mathematical hypotheses used to define the class of artifacts to be synthesized within the overall verification or synthesis problem. The sciductive approach constrains inductive and deductive reasoning using structure hypotheses, and actively combines inductive and deductive reasoning: for instance, deductive techniques generate examples for learning, and inductive reasoning is used to guide the deductive engines. We describe the methodology in detail, with comparison to related work, in Section 2.

Let us reflect on the combined use of induction and deduction for verification and synthesis. Automatic techniques for verification and synthesis are typically deductive. However, one can argue that humans often employ a combination of inductive and deductive reasoning while performing verification or synthesis. For example, while proving a theorem, one often starts by working out examples and trying to find a pattern in the properties satisfied by those examples. The latter step is a process of inductive generalization. These patterns might take the form of lemmas or background facts that then guide a deductive process of proving the statement of the theorem from known facts and previously established theorems (rules). Similarly, while creating a new design, one often starts by enumerating sample behaviors that the design must satisfy and hypothesizing components that might be useful in the design process; one then systematically combines these components, using design rules, to obtain a candidate artifact. The process usually iterates between inductive and deductive reasoning until the final artifact is obtained. It is this combination of inductive and deductive reasoning that we seek to formalize using the notion of sciduction.

We demonstrate the sciductive approach to automatic verification and synthesis using three applications:

1. Timing analysis of software (Section 3);
2. Synthesis of loop-free programs (Section 4), and
3. Synthesizing switching logic for control of hybrid systems (Section 5).

The first application of the sciduction approach addresses the problem of environment modeling; the second, the problem of inadequate specifications, and the third tackles the problem of high computational complexity. Some future applications are also explored in Section 6.

## 2 Sciduction: Formalization and Related Work

We begin with a formalization of the sciductive approach, and then compare it to related work. This section assumes some familiarity with basic terminology in formal verification and machine learning — see the relevant books by Clarke et al. [14], Manna and Pnueli [34], and Mitchell [38] for an introduction.

### 2.1 Verification and Synthesis Problems

As discussed in Section 1, an instance of a verification problem is defined by a triple $\langle S, E, \Phi \rangle$, where $S$ denotes the system, $E$ is the environment, and $\Phi$ is the property to be verified. Here we assume that $S$, $E$, and $\Phi$ are described formally, in mathematical notation. Similarly, an instance of a synthesis problem is defined by the pair $\langle E, \Phi \rangle$, where the symbols have the same meaning. In both cases, as noted earlier, it is possible *in practice* for the descriptions of $S$, $E$, or $\Phi$ to be missing or incomplete; in such cases, the missing components must be synthesized as part of the overall verification or synthesis process.

A *family of verification or synthesis problems* is a triple $\langle C_S, C_E, C_\Phi \rangle$ where $C_S$ is a formal description of a class of systems, $C_E$ is a formal description of a class of environment models, and $C_\Phi$ is a formal description of a class of specifications. In the case of synthesis, $C_S$ defines the class of systems to be synthesized.

## 2.2 Elements of Sciduction

An instance of sciduction can be described using a triple $\langle \mathcal{H}, I, \mathcal{D} \rangle$, where the three elements are as follows:

1. *A structure hypothesis*, $\mathcal{H}$, encodes our hypothesis about the form of the *artifact to be synthesized*, whether it be an environment model, an inductive invariant, a program, or a control algorithm (or any portion thereof);

2. *An inductive inference engine*, $I$, is an algorithm for *learning from examples* an artifact $h$ defined by $\mathcal{H}$, and

3. A *deductive engine*, $\mathcal{D}$, is a *lightweight decision procedure* that applies deductive reasoning to answer queries generated in the synthesis or verification process.

We elaborate on these elements below. The context of synthesis is used to explain the central ideas in the sciductive approach. While the application of these ideas to verification is symmetric, we will note points specific to verification or synthesis as they arise.

### 2.2.1 Structure Hypothesis

*The structure hypothesis, $\mathcal{H}$, encodes our hypothesis about the form of the artifact to be synthesized.*

Formally $\mathcal{H}$ is a (possibly infinite) set of *artifacts*. $\mathcal{H}$ encodes a hypothesis that the system to be synthesized falls in a subclass $C_\mathcal{H}$ of $C_S$ (i.e., $C_\mathcal{H} \subseteq C_S$). Note that $\mathcal{H}$ needs not be the same as $C_\mathcal{H}$, since the artifact being synthesized might just be a portion of the full system description, such as the guard on transitions of a state machine, or the assignments to certain variables in a program. Each artifact $h \in \mathcal{H}$, in turn, corresponds to a unique set of *primitive elements* that defines its semantics. The form of the primitive element depends on the artifact to be synthesized.

More concretely, here are two examples of a structure hypothesis $\mathcal{H}$:

1. Suppose that $C_S$ is the set of all finite automata over a set of input variables $V$ and output variables $U$ satisfying a specification $\Phi$. Consider the structure hypothesis $\mathcal{H}$ that restricts the finite automata to be synchronous compositions of automata from a finite library $L$. The artifact to be synthesized is the entire finite automaton, and so, in this case, $\mathcal{H} = C_\mathcal{H}$. Each element $h \in \mathcal{H}$ is one such composition of automata from $L$. A primitive element is an input-output trace in the language of the finite automaton $h$.

2. Suppose that $C_S$ is the set of all hybrid automata [1], where the guards on transitions between modes can be any region in $\mathbb{R}^n$ but where the modes of the automaton are fixed. A structure hypothesis $\mathcal{H}$ can restrict the guards to be hyperboxes in $\mathbb{R}^n$ — i.e., conjunctions of upper and lower bounds on continuous state variables. Each $h \in \mathcal{H}$ is one such hyperbox, and a primitive element is a point in $h$. Notice that $\mathcal{H}$ defines a subclass of hybrid automata $C_\mathcal{H} \subset C_S$ where the guards are $n$-dimensional hyperboxes. Note also that $\mathcal{H} \neq C_\mathcal{H}$ in this case.

A structure hypothesis $\mathcal{H}$ can be syntactically described in several ways. For instance, in the second example above, $\mathcal{H}$ can define a guard either set-theoretically as a hyperbox in $\mathbb{R}^n$ or using mathematical logic as a conjunction of atomic formulas, each of which is an interval constraint over a real-valued variable.

### 2.2.2 Inductive Inference

*The inductive inference procedure, I, is an algorithm for learning from examples an artifact $h \in \mathcal{H}$.*

While any inductive inference engine can be used, in the context of verification and synthesis we expect that the learning algorithms $I$ have one or more of the following characteristics:

- $I$ performs *active learning*, selecting the examples that it learns from.
- Examples and/or labels for examples are generated by one or more *oracles*. The oracles could be implemented using deductive procedures or by evaluation/execution of a model on a concrete input. In some cases, an oracle could be a human user.
- A deductive procedure is invoked in order to synthesize a concept (artifact) that is consistent with a set of labeled examples. The idea is to formulate this synthesis problem as a decision problem where the concept to be output is generated from the satisfying assignment.

### 2.2.3 Deductive Reasoning

*The deductive engine, $\mathcal{D}$, is a lightweight decision procedure that applies deductive reasoning to answer queries generated in the synthesis or verification process.*

The word "lightweight" refers to the fact that the decision problem being solved by $\mathcal{D}$ must be easier, in theoretical or practical terms, than that corresponding to the overall synthesis or verification problem.

In theoretical terms, "lightweight" means that at least one of the following conditions must hold:

1. $\mathcal{D}$ must solve a problem that is a strict special case of the original.
2. One of the following two cases must hold:
   (i) If the original (synthesis or verification) problem is decidable, and the worst-case running time of the best known procedure for the original problem is $O(T(n))$, then $\mathcal{D}$ must run in time $o(T(n))$.
   (ii) If the original (synthesis or verification) problem is undecidable, $\mathcal{D}$ must solve a decidable problem.

In practical terms, the notion of "lightweight" is fuzzier: intuitively, the class of problems addressed by $\mathcal{D}$ must be "more easily solved in practice" than the original problem class. For example, $\mathcal{D}$ could be a finite-state model checker that is invoked only on abstractions of the original system produced by, say, localization abstraction [28] — it is lightweight if the abstractions are solved faster in practice than the original concrete instance. Due to this fuzziness, it is preferable to define "lightweight" in theoretical terms whenever possible.

$\mathcal{D}$ can be used to answer queries generated by $I$, where the query is typically formulated as a decision problem to be solved by $\mathcal{D}$. Here are some examples of tasks $\mathcal{D}$ can perform and the corresponding decision problems:

- Generating examples for the learning algorithm.
  *Decision problem:* "does there exist an example satisfying the criterion of the learning algorithm?"
- Generating labels for examples selected by the learning algorithm.
  *Decision problem:* "is $L$ the label of this example?"
- Synthesizing candidate artifacts.
  *Decision problem:* "does there exists an artifact consistent with the observed behaviors/examples?"

### 2.2.4 Discussion

We now make a few remarks on the formalization of sciduction introduced above.

In the above description of the structure hypothesis, $\mathcal{H}$ only "loosely" restricts the class of systems to be synthesized, allowing the possibility that $\mathcal{C}_{\mathcal{H}} = \mathcal{C}_S$. We argue that a tighter restriction is often desirable. One important role of the structure hypothesis is to reduce the search space for synthesis, by restricting the class of artifacts $\mathcal{C}_S$. For example, a structure hypothesis could be a way of codifying the form of human insight to be provided to the synthesis process. Additionally, restricting $C_\mathcal{H}$ also aids in inductive inference. Fundamentally, the effectiveness of inductive inference (i.e., of $I$) is limited by the examples presented to it as input; therefore, it is important not only to select examples carefully, but also for the inference to generalize well beyond the presented examples. For this purpose, the structure hypothesis should place a strict restriction on the search space, by which we mean that $C_\mathcal{H} \subsetneq C_S$. The justification for this stricter restriction comes from the importance of *inductive bias* in machine learning. Inductive bias is the set of assumptions required to *deductively* infer a concept from the inputs to the learning algorithm [38]. If one places no restriction on the type of systems to be synthesized, the inductive inference engine $I$ is unbiased; however, an unbiased learner will learn an artifact that is consistent only with the provided examples, with no generalization to unseen examples. As Mitchell [38] writes: "a learner that makes no a priori assumptions regarding the identity of the target concept has no rational basis for classifying any unseen instances." Given all these reasons, it is desirable for the structure hypothesis $\mathcal{H}$ to be such that $C_\mathcal{H} \subsetneq C_S$. This is the case in the three applications demonstrated in Sections 3, 4, and 5.

Another point to note is that it is possible to use randomization in implementing $I$ and $\mathcal{D}$. For example, a deductive decision procedure that uses randomization can generate a YES/NO answer with high probability.

Next, although we have defined sciduction as combining a single inductive engine with a single deductive engine, this is only for simplicity of the definition and poses no fundamental restriction. One can always view multiple inductive (deductive) engines as a being contained in a single inductive (deductive) procedure where this outer procedure passes its input to the appropriate "sub-engine" based on the type of input query.

Finally, in our definition of sciduction, we do not advocate any particular technique of combining inductive and deductive reasoning. Indeed, we envisage that there are many ways to "configure" the combination of $\mathcal{H}$, $\mathcal{D}$, and $I$, perhaps using inductive procedures within deductive engines and vice-versa. Any mode of integrating $\mathcal{H}$, $I$, and $\mathcal{D}$ that satisfies the requirements stated above on each of those three elements is admissible. We expect that the particular requirements of each application will define the mode of integration that works best for that application. We present illustrative examples in Sections 3, 4, and 5.

In Section 2.4, we discuss several examples of related work, to differentiate the sciductive approach from other methods.

## 2.3 Soundness and Completeness Guarantees

It is highly desirable for verification or synthesis procedures to provide *soundness* and *completeness* guarantees. In this section, we discuss the form these guarantees take for a procedure based on sciduction.

A verifier is said to be *sound* if, given an arbitrary problem instance $\langle S, E, \Phi \rangle$, the verifier outputs "YES" only if $S \| E \models \Phi$. The verifier is said to be *complete* if it outputs "NO" when $S \| E \not\models \Phi$.

The definitions for synthesis are similar. A synthesis technique is *sound* if, given an arbitrary problem instance $\langle E, \Phi \rangle$, if it outputs $S$, then $S \| E \models \Phi$. A synthesis technique is *complete* if, when there exists $S$ such that $S \| E \models \Phi$, it outputs at least one such $S$.

Formally, for a verification/synthesis procedure $\mathcal{P}$, we denote the statement "$\mathcal{P}$ is sound" by $\texttt{sound}(\mathcal{P})$.

Note that we can have probabilistic analogs of soundness and completeness. Informally, a verifier is *probabilistically sound* if it is sound with "high probability;" we will leave a more precise discussion of this point to a later stage in this paper when it becomes relevant. In this section, we also use $\texttt{sound}(\mathcal{P})$ to denote probabilistic soundness.

### 2.3.1 Validity of the Structure Hypotheses

In sciduction, the existence of soundness and completeness guarantees depends on the validity of the structure hypothesis. Informally, we say that the structure hypothesis $\mathcal{H}$ is *valid* if the artifact to be synthesized, if one exists, is guaranteed to be an element of the class $C_\mathcal{H}$.

Let us elaborate on what we mean by the phrase "the artifact to be synthesized":

- In the context of a synthesis problem, this is relatively easy: one seeks an element $c$ of $C_S$ that satisfies a specification $\Phi$. If $\Phi$ is available as a formal specification, the above phrase is precisely defined. However, as noted earlier, one of the challenges with synthesis can be the absence of good formal specifications. In such cases, we use $\Phi$ to denote a "golden" specification that one would have in the ideal scenario.
- For verification, there can be many artifacts to be synthesized, such as inductive invariants, abstractions, or environment assumptions. Each such artifact is an element of a set $C_S$. The "specification" for each such synthesis "sub-task", generating a different kind of artifact, is different. For invariant generation, $C_S$ is the set of candidate invariants, and the specification is that the artifact $c \in C_S$ be an inductive invariant of the system $S$. For abstractions, $C_S$ defines the set of abstractions, and the specification is that $c \in C_S$ must be a sound *and* precise abstraction with respect to the property to be verified, $\Phi$; here "precise" means that no spurious counterexamples will be generated. We will use $\Psi$ to denote the cumulative specification for all synthesis sub-tasks in the verification problem.

Thus, for both verification and synthesis, the existence of an artifact to be synthesized can be expressed as the following logical formula:
$$\exists c \in C_S \,.\, c \models \Psi$$
where, for synthesis, $\Psi = \Phi$, and, for verification, $\Psi$ denotes the cumulative specification for the synthesis sub-tasks, as discussed above.

Similarly, the existence of an artifact to be synthesized that additionally satisfies the structure hypothesis $\mathcal{H}$ is written as:
$$\exists c \in C_{\mathcal{H}} \,.\, c \models \Psi$$

Given the above logical formulas, we define the statement "the structure hypothesis is *valid*" as the validity of the logical formula $\texttt{valid}(\mathcal{H})$ given below:
$$\texttt{valid}(\mathcal{H}) \triangleq (\exists c \in C_S \,.\, c \models \Psi) \implies (\exists c \in C_{\mathcal{H}} \,.\, c \models \Psi) \quad (1)$$

In other words, if there exists an artifact to be synthesized (that satisfies the corresponding specification $\Psi$), then there exists one satisfying the structure hypothesis.

Note that $\texttt{valid}(\mathcal{H})$ is trivially valid if $C_{\mathcal{H}} = C_S$. Indeed, one extremely effective technique in verification, counterexample-guided abstraction refinement (CEGAR), can be seen as a form of sciduction where typically $C_{\mathcal{H}} = C_S$. (Sec. 2.4 has a more detailed discussion of the link between CEGAR and sciduction.) However, in some cases, $\texttt{valid}(\mathcal{H})$ can be proved valid even without $C_{\mathcal{H}} = C_S$; see Sec. 5 for an example.

### 2.3.2 Conditional Soundness

A verification/synthesis procedure following the sciduction paradigm must satisfy a conditional soundness guarantee: procedure $\mathcal{P}$ must be *sound* (or probabilistically sound) *if the structure hypothesis is valid*.

Without such a requirement, the sciductive procedure is a heuristic, best-effort verification or synthesis procedure. (It could be extremely useful, nonetheless.) With this requirement, we have a mechanism to formalize the assumptions under which we obtain soundness — namely, the structure hypothesis.

More formally, the soundess requirement for a sciductive procedure $\mathcal{P}$ can be expressed as the following logical expression:
$$\texttt{valid}(\mathcal{H}) \implies \texttt{sound}(\mathcal{P}) \quad (2)$$

Note that one must prove $\texttt{sound}(\mathcal{P})$ under the assumption $\texttt{valid}(\mathcal{H})$, just like one proves unconditional soundness. The point is that making a structure hypothesis can allow one to devise procedures and prove soundness where previously this was difficult or impossible.

Where completeness is also desirable, one can formulate a similar notion of conditional completeness. We will mainly focus on soundness in this paper, since in our experience it is the more valuable property.

## 2.4 Comparison with Related Work

In both ancient and modern philosophy, there is a long history of arguments about the distinction between induction and deduction and their relationship and relative importance. This literature, although very interesting, is not directly relevant to the discussion in this paper.

Within computer science and engineering, the field of artificial intelligence (AI) has long studied inductive and deductive reasoning and their connections (see, e.g., [49]). As mentioned earlier, Mitchell [38] describes how inductive inference can be formulated as a deduction problem where inductive bias is provided as an additional input to the deductive engine. *Inductive logic programming* [40], an approach to machine learning, blends induction and deduction by performing inference in first-order theories using examples and background knowledge. Combinations of inductive and deductive reasoning have also been explored for synthesizing programs (plans) in AI; for example, the SSGP approach [19] generates plans by sampling examples, generalizing from those samples, and then proving correctness of the generalization.

Our focus is on the use of combined inductive and deductive reasoning in *formal verification and synthesis*. While several techniques for verification and synthesis combine subsets of induction, deduction, and structure hypotheses, there are important distinctions between many of these and the sciduction approach. We present below a representative sample of related work.

### 2.4.1 Closely Related Work

We first survey prior work in verification and synthesis that has provided inspiration for formulating the sciductive approach. We note that many of these prior techniques can be formulated as instances of sciduction. Indeed, sciduction can be seen as a "lens" through which one can view the common ideas amongst these techniques so as to extend and apply them to new problem domains.

**Counterexample-Guided Abstraction Refinement (CEGAR).** In CEGAR [13], depicted in Fig. 3, an abstract model is synthesized so as to eliminate spurious counterexamples. CEGAR solves a synthesis sub-task of generating abstract models that are *sound* (they contain all behaviors of the original system) and *precise* (any counterexample for the abstract model is also a counterexample for the original system). The synthesized artifact is thus the abstract model. One can view CEGAR as an instance of sciduction as follows:

- The *abstract domain*, which defines the form of the abstraction function and hence the abstract model, is the structure hypothesis. For example, in verifying digital circuits, one might use localization abstraction [28], in which abstract states are cubes over the state variables.
- The inductive engine $I$ is an algorithm to learn a new abstraction function from a spurious counterexample. Consider the case of localization abstraction. The traditional approach in CEGAR is to walk the lattice of abstraction functions, from most abstract (hide all variables) to least abstract (the original system); this is an instance of learning based on version spaces [38]. However, alternative learning algorithms (such as induction on decision trees) can also be used, as demonstrated by Gupta [22].
- The deductive engine $\mathcal{D}$, for finite-state model checking, comprises the model checker and a SAT solver. The model checker is invoked on the abstract model to check the property of interest, while the SAT solver is used to check if a counterexample is spurious.

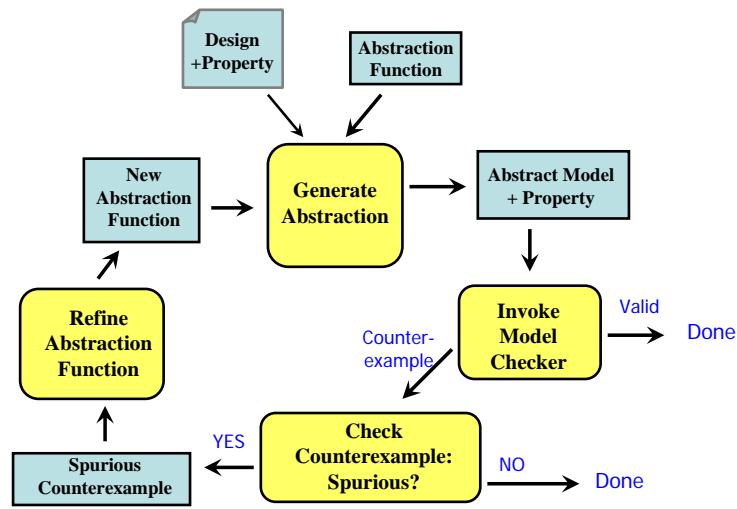

Figure 3: Counterexample-guided abstraction refinement (CEGAR).

As observed earlier, since the original system is a valid abstract model, $\mathcal{C}_\mathcal{H} = \mathcal{C}_S$ in the case of CEGAR. Thus, the structure hypothesis is valid, and the notion of soundness reduces to the traditional notion.

**Learning for Assume-Guarantee Reasoning and Compositional Verification.** The use of learning algorithms has been investigated extensively in the context of synthesizing environment assumptions for compositional verification. Most of these techniques are based on Angluin's $L^*$ algorithm and its variants; see [16] for a recent collection of papers on this topic. These techniques are an instance of sciduction, similar to CEGAR, in which typically no restrictive structure hypothesis is made (i.e., $\mathcal{C}_\mathcal{H} = \mathcal{C}_S$). For example, for techniques that target finite-state model checking, the synthesized environment assumptions can be any finite-state machine that interfaces with the system. The $L^*$ algorithm is a learning algorithm based on queries and counterexamples. The counterexamples are generated by the model checker, which forms the deductive procedure in this case. One possible restriction on the structure hypothesis would be to limit the input or output alphabets for the learning algorithm.

**Invariant Generation.** One of the important steps in verification based on model checking or theorem proving is the construction of *inductive invariants*. (Here "inductive" refers to the use of mathematical induction.) One often needs to strengthen the main safety property with auxiliary inductive invariants so as to succeed at proving/disproving the property.

In recent years, an effective approach to generating inductive invariants is to assume that they have a particular structural form, use simulation/testing to prune out candidates, and then use a SAT/SMT solver or model checker to prove those candidates that remain. This is an instance of sciduction, and can be very effective. For example, these strategies are implemented in the ABC verification and synthesis system [8] and described in part in Michael Case's PhD thesis [11]. The structure hypothesis $\mathcal{H}$ defines the space of candidate invariants as being either constants (literals), equivalences, implications, or in some cases, random clauses or based on $k$-cuts in the and-inverter graph. The inductive inference engine is very rudimentary: it just keeps all instances of invariants that match $\mathcal{H}$ and are consistent with simulation traces. The deductive engine is a SAT solver. Clearly, in this case, the structure hypothesis is restrictive in that the procedure does not seek to find arbitrary forms of invariants. However, the verification procedure is still sound, because if a suitable inductive invariant is not found, one may fail to prove the property, but a buggy system will not be deemed correct.

This idea has also been explored in software verification, by combining the Daikon system [17] for generating likely program invariants from traces with deductive verification systems such as ESC/Java [18].

**Software Synthesis by Sketching.** Programming by sketching is a novel approach to synthesizing software by encoding programmer insight in the form of a *partial program*, or "sketch" [56, 57, 55]. One of the main algorithmic approaches to sketching can be viewed as an instance of sciduction. The partial program imposes a structure hypothesis on the synthesis problem. An algorithmic approach central to this work is counterexample-guided inductive synthesis (CEGIS) [57, 55], which operates in a manner similar to CEGAR. The inductive and deductive procedures are just as those in CEGAR. Since CEGIS repeatedly calls a verifier in its main loop, it is effective for systems where the verification problem is significantly easier than the synthesis problem for the same class of systems and specifications. For example, in the synthesis of finite (bit-vector) loop-free programs [57], the verifier solves SAT, whereas the synthesis problem is expressible as a 2-QBF problem (satisfiability of quantified Boolean formulas with one alternation).

**Discussion.** Techniques such as CEGAR, CEGIS, learning-based compositional verification, and dynamic invariant generation have proved very successful and influential in their respective domains. However, the reader may note that the problems addressed in Sections 3, 4, and 5 are all problems that are difficult to tackle using techniques such as CEGIS or CEGAR for the reasons described in Section 1 — the lack of a precise specification or environment model, or the high computational complexity of verification that makes it difficult to repeatedly invoke a verifier within a synthesis loop. Our goal at formalizing the notion of sciduction is to provide a common framework to build upon existing successes, such as CEGAR, and go beyond the problems addressed by these methods.

### 2.4.2 Other Related Work

We now highlight some other related work and their distinctions with sciduction.

**Verification-Driven Synthesis.** Srivastava et al. [59] have proposed a verification-driven approach to synthesis (called VS3), where programs with loops can be synthesized from a scaffold comprising of a logical specification of program functionality, and domain constraints and templates restricting the space of synthesizable programs. The latter is a structure hypothesis. However, the approach is not sciduction since the synthesis techniques employed are purely deductive in nature. More recently, Srivastava et al. [58] have proposed a path-based inductive synthesis approach. An inverse of a program is synthesized by exploring a carefully-chosen subset of that program's paths — the inductive generalization is that the program synthesized to yield the inverse for the chosen subset of paths is also deemed to yield the correct inverse for all other paths. In contrast with sciduction, there is no guarantee that under a valid structure hypothesis, the synthesis routine will yield the correct program, if one exists.

**Boolean Satisfiability (SAT) Solving.** Modern SAT solvers use the *conflict-driven clause learning* (CDCL) approach [31]. In a CDCL SAT solver, whenever the solver explores a partial assignment that leads to a conflict (i.e., the formula evaluates to false), it "learns" a new clause that captures a reason for that conflict. This learned clause can exclude not only this falsifying partial assignment, but potentially also other similar assignments. Thus, one can view CDCL SAT solving as using a subroutine (clause learning) that generalizes from experiments, each experiment being a partial assignment explored by the SAT solver. However, note that the form of generalization employed is not inductive. Clause learning is, in fact, a *deductive procedure*: using the clause database as background facts, a particular CDCL clause learning strategy can be seen as a way of applying the resolution proof rule to clauses involved in implications generated by the falsifying partial assignment. Additionally, no structure hypothesis is made. Therefore, although there appear to be similarities with sciduction, the CDCL approach to SAT solving also has important differences.

**Program Analysis Using Relevance Heuristics.** McMillan [37] describes the idea of verification based on "relevance heuristics", which is the notion that facts useful in proving special cases of the verification problem are likely to be useful in general. This idea is motivated by the similar approach taken in (CDCL) SAT solvers.

A concrete instance of this approach is interpolation-based model checking [36], where a proof of a special case (e.g., the lack of an assertion failure down a certain program path) is used to generate facts relevant to solving the general verification problem (e.g., correctness along all program paths). Although this work generalizes from special cases, the generalization is not inductive, and no structure hypothesis is involved.

**Automata-Theoretic Synthesis from Linear Temporal Logic (LTL).** One of the classic approaches to synthesis is the automata-theoretic approach for synthesizing a finite-state transducer (FST) from an LTL specification, pioneered by Pnueli and Rosner [42]. The approach is a purely deductive one, with a final step that involves solving an emptiness problem for tree automata. No structure hypothesis is made on the FST being synthesized. Although advances have been made in the area of synthesis from LTL, for example in special cases [41], some major challenges remain: (i) writing complete specifications is tedious and error-prone, and (ii) the computational complexity for general LTL is doubly-exponential in the size of the specification. It would be interesting to explore if inductive techniques can be combined with existing deductive automata-theoretic procedures to form an effective sciductive approach to some class of systems or specifications.

## 3 Quantitative Analysis of Sofware

The analysis of quantitative properties, such as bounds on timing and power, is central to the design of reliable embedded software and systems, such as those in automotive, avionics, and medical applications. Fundamentally, such properties depend not only on program logic, but also on details of the program's environment. The environment includes many things — the processor, characteristics of the memory hierarchy, the operating system, the network, etc. Moreover, in contrast with many other verification problems, the environment must be modeled with a relatively high degree of precision. Most state-of-the-art approaches to worst-case execution time (WCET) analysis require significant manual modeling, which can be tedious, error-prone and time consuming, even taking several months to create a model of a relatively simple microcontroller. See [50] for a more detailed description of the challenges in quantitative analysis of software.

### 3.1 The Problem

For simplicity, we will consider the following representative timing analysis problem:

⟨TA⟩ Given a terminating program $P$, its platform (environment) $E$, and a fixed starting state of $E$, is the execution time of $P$ on $E$ always at most $\tau$?

If the execution time can exceed $\tau$, it is desirable to obtain a test case (comprising a state of $P$) that shows how the bound of $\tau$ is exceeded.

The main challenge in solving this problem, as noted earlier, is the generation of a model of the environment $E$. We illustrate this challenge briefly using a toy example. The complexity of the timing analysis arises from two dimensions of the problem: the *path dimension*, where one must find the right computation path in the task, and the *state dimension*, where one must find the right (starting) environment state to run the task from. Moreover, these two dimensions interact closely; for example, the choice of path can affect the impact of the starting environment state.

Consider the toy C program in Fig. 4(a). It contains a loop, which executes at most once. Thus, the control-flow graph (CFG) of the program can be unrolled into a directed acyclic graph (DAG), as shown in Fig. 4(b). Suppose we execute this program on a simple processor with an in-order pipeline and a data cache. Consider executing this program from the state where the cache is empty. The final statement of the program, `*x += 2`, contains a load, a store, and an arithmetic operation. If the left-hand path is taken, the load will suffer a cache miss; however, if the right-hand path is taken, there is a cache hit. The difference in timing between a cache hit and a miss can be an order of magnitude. Thus, the time taken by this statement depends on the program path taken. However, if the program were executed from a state with the data in the cache, there will be a cache hit even if the left-hand path is taken.

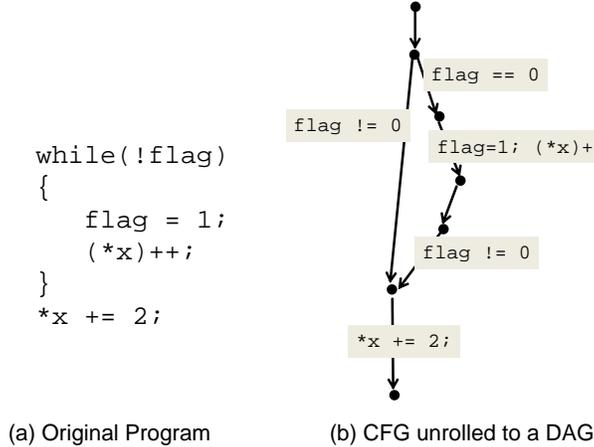

Figure 4: **Simple Illustrative Example**

Thus, even with this toy program and a simple processor, one can observe that a timing analysis tool must explore the space of all possible program paths and all possible environment states – both potentially exponentially-large search spaces. Most WCET tools assume a known starting environment state, and attempt to predict worst-case timing by inferring worst-case environment states at basic block boundaries. For this, they must (i) model the platform precisely to predict timing at basic block boundaries, and (ii) search an exponentially-large environment state space at these boundaries. If the starting environment state is unknown, the problem is even harder!

We show in the next section that sciduction offers a promising approach to address this challenge of environment modeling.

## 3.2 Our Approach: GAMETIME

Automatic inductive inference of models offers a way to mitigate the challenge of environment modeling. In our approach, termed GAMETIME, a *program-specific timing model* of the platform is inferred from observations of the program's timing that are automatically and systematically generated. The program-specificity is an important difference from traditional approaches, which seek to manually construct a timing model that works for *all* programs one might run on the platform. GAMETIME only requires one to run end-to-end measurements on the target platform, making it easy to port to new platforms.

The central idea in GAMETIME is to view the platform as an adversary that controls the choice and evolution of the environment state, while the tool has control of the program path space. The problem is then a formulated as a game between the tool and the platform. GAMETIME uses a sciductive approach to solve this game based on the following elements:

- *Structure hypothesis:* The platform $E$ is modeled as an adversarial process that selects weights on the edges of the control-flow graph of the program $P$ in two steps: first, it selects the path-independent weights $w$, and then the path-dependent component $\pi$. Formally, $w$ cannot depend on the program path being executed, whereas $\pi$ is drawn from a distribution which is a function of that path. Both $w$ and $\pi$ are elements of $\mathbb{R}^m$, where $m$ is the number of edges in the CFG after unrolling loops and inlining function calls. We term $w$ as the *weight* and $\pi$ as the *perturbation*, and the structure hypothesis as the *weight-perturbation model*. More specifically, one structure hypothesis $\mathcal{H}$ used by GAMETIME is to define a space of environment models as a set of processes that select a pair $(w, \pi)$ every time the program $P$ runs, where additionally the pair satisfies the following constraints (see [54] for details):

1. The mean perturbation along any path is bounded by a quantity $\mu_{\max}$.
2. (for worst-case analysis) The worst-case path is the unique longest path by a specified margin $\rho$.

- *Inductive inference:* The inductive inference routine is a learning algorithm that operates in a *game-theoretic online setting*. The task of this algorithm is to learn the $(w, \pi)$ model from measurements. The idea in GAMETIME is to measure execution times of $P$ along so-called *basis paths*, choosing amongst these basis paths uniformly at random over a number of trials. A $(w, \pi)$ model is inferred from the end-to-end measurements of program timing along each of the basis paths. See [54] for further details.
- *Deductive reasoning:* An SMT solver forms the deductive engine for GAMETIME. The basis paths constitute the examples for the inductive learning algorithm. These examples are generated using SMT-based test generation — from each candidate basis path, an SMT formula is generated such that the formula is satisfiable iff the path is feasible.

The GAMETIME approach, along with an exposition of theoretical and experimental results, including comparisons with other methods, is described in existing papers [53, 54, 52]. We only give a brief overview here.

Figure 5 depicts the operation of GAMETIME. As shown in the top-left corner, the process begins with the generation of the control-flow graph (CFG) corresponding to the program, where all loops have been unrolled to a maximum iteration bound, and all function calls have been inlined into the top-level function. The CFG is assumed to have a single source node (entry point) and a single sink node (exit point); if not, dummy source and sink nodes are added. The next step is a critical one, where a subset of program paths, called *basis paths* are extracted. These basis paths are those that form a basis for the set of all paths, in the standard linear algebra sense of a basis. A *satisfiability modulo theories (SMT) solver* — the deductive engine — is invoked to ensure that the generated basis paths are feasible. For each feasible basis path generated, the SMT solver generates a test case that drives program execution down that path. Thus a set of *feasible basis paths* is generated that spans the entire space of feasible program paths, along with the corresponding test cases.

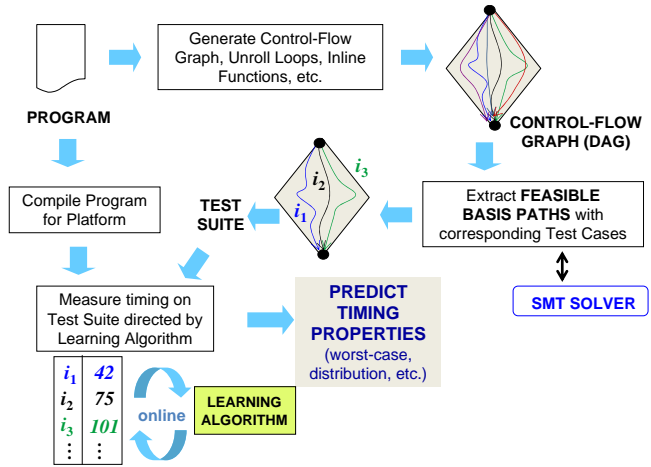

Figure 5: GAMETIME **overview**

The program is then compiled for the target platform, and executed on these test cases. In the basic GAMETIME algorithm (described in [53, 54]), the sequence of tests is randomized, with basis paths being chosen uniformly at random to be executed. The overall execution time of the program is recorded for each test case. From these end-to-end execution time measurements, GAMETIME's learning algorithm generates the $(w, \pi)$ model that can then be used for timing analysis. In contrast with most existing tools for timing analysis (see, e.g., [46]), GAMETIME can not only be used for WCET estimation, it can also be used to predict execution time of arbitrary program paths, and certain execution time statistics (e.g., the distribution of times). For example, to answer the problem $\langle TA \rangle$ presented in the preceding section, GAMETIME would predict the longest path, execute it to compute the corresponding timing $\tau^*$, and compare that time with $\tau$: if $\tau^* \leq \tau$, then GAMETIME returns "YES", otherwise it returns "NO" along with the corresponding test case.

### 3.3 Guarantees and Results

Assuming the structure hypothesis holds, GAMETIME answers the timing analysis question $\langle TA \rangle$ with high probability. In other words, if the structure hypothesis is valid, GAMETIME is *probabilistically sound and complete* in the following sense:

> Given any $\delta > 0$, if one runs a number of tests that is polynomial in $\ln \frac{1}{\delta}$, $\mu_{max}$, and the program parameters, GAMETIME will report the correct YES/NO answer to Problem $\langle TA \rangle$ with probability at least $1 - \delta$.

See the theorems in [53, 54] for details.

Experimental results indicate that in practice GAMETIME can accurately predict not only the worst-case path (and thus the WCET) but also the distribution of execution times of a task from various starting environment states. As a sample result, we used GAMETIME to estimate the distribution of execution times of a modular exponentiation function function for an 8-bit exponent (256 program paths) by testing only (9) basis paths. The experiments were performed for the StrongARM-1100 processor – which implements the ARM instruction set with a 5-stage pipeline and both data and instruction caches – using the SimIt-ARM cycle-accurate simulator [44]. Fig. 6 shows the predicted and actual distribution of execution times – we see that GAMETIME predicts the distribution perfectly. Also, GAMETIME correctly predicts the WCET (and produces the corresponding test case: the 8-bit exponent is 255).

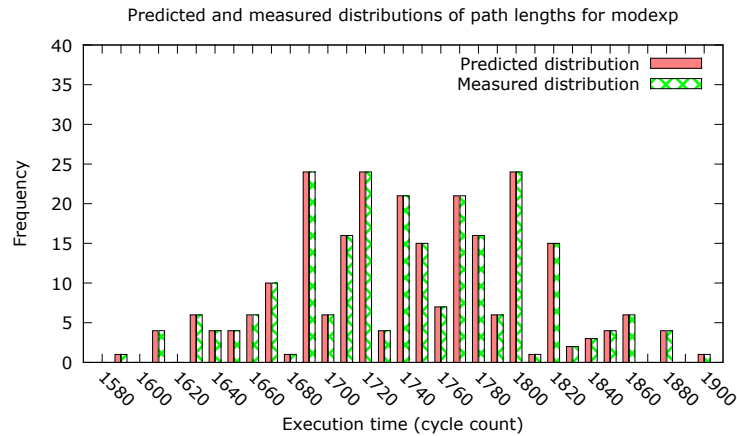

Figure 6: **Actual (striped) and Predicted (shaded) Execution Times for Modular Exponentiation Code.**

# 4 Component-Based Program Synthesis

Automatic synthesis of programs has long been one of the holy grails of software engineering. In this section, we focus on a particular application of synthesis to *program understanding* — viz., *program deobfuscation* with a focus on malicious programs. The need for deobfuscation techniques has arisen in recent years, especially due to an increase in the amount of malicious code (malware) that tends to be obfuscated [60]. Currently, human experts use decompilers and manually reverse engineer the resulting code (see, e.g., [43]). Clearly, this is a tedious task that could benefit from automated tool support.

## 4.1 The Problem

Our idea is to view the malware deobfuscation problem as a program (re-)synthesis problem [23]. The focus is on re-synthesizing fragments of the original malicious program where each such fragment, though it may contain loops, is essentially equivalent to a loop-free program. The main challenge for this synthesis problem is the lack of a good formal specification — in fact, the only available "specification" is the obfuscated malicious program itself. This is not an ideal specification to work with; not only might it contain maliciously-inserted constructs that make static analysis and satisfiability solving hard, its translation to a logical specification is likely to generate a large, complex formula.

Our approach to this problem is to view the obfuscated program as an *I/O oracle* that maps a given program input (starting state) to the desired output (ending state).[2] The problem is then to synthesize the program using a small number of queries to the I/O oracle.

An important advantage of this I/O oracle view is that the complexity of the synthesis procedure (e.g., the number of queries to the I/O oracle) required is a function of the intrinsic functionality of the program, not of the syntactic obfuscations applied to it.

## 4.2 Our Approach

Our sciductive approach to this synthesis problem has the following ingredients:

- *Structure Hypothesis:* Programs are assumed to be loop-free compositions of components drawn from a finite component library *L*. Each component in this library implements a programming construct that is essentially a *bit-vector circuit* — the outputs are bit-vector functions of a set of input bit-vectors. Conditional statements are allowed, but loops are disallowed. Thus, $\mathcal{C}_\mathcal{H}$ is the set of all programs that can be synthesized as syntactically legal compositions of components from *L*.
- *Inductive Inference:* The inductive inference routine is an algorithm that learns a program from *distinguishing inputs* — those inputs that can distinguish between non-equivalent programs in $\mathcal{C}_\mathcal{H}$ which are consistent with past interaction with the I/O oracle. The inductive routine starts with one or more randomly chosen inputs and their outputs obtained from the I/O oracle. On each iteration, the routine constructs an SMT formula whose satisfying assignment yields a program consistent with all input-output examples seen so far. It also queries the SMT solver for another such program which is semantically different from the first, as well as a distinguishing input that demonstrates this semantic difference. If no such alternative program exists, the process terminates. Otherwise, the process is repeated until a semantically unique program is obtained.

  Our algorithm is motivated by the characterization of the *optimal teaching sequence* by Goldman and Kearns [20]. In that paper, the authors introduce the concept of *teaching dimension* of a concept class as the minimum number of examples a teacher (oracle) must reveal to uniquely identify *any* target concept from that class. They show that the generation of an *optimal teaching sequence* of examples is equivalent to a minimum set cover problem. In the set cover problem for a given target concept, the universe of elements is the set of all incorrect concepts (programs) and each set $S_i$, corresponding to example $x_i$, contains concepts that are differentiated from the target concept by this example $x_i$. Our algorithm computes such a distinguishing example in each iteration, and terminates when it has computed a "set cover" that distinguishes the target concept from all other candidate concepts (the "universe"). In practice, our algorithm has required only a small number of iterations, indicating that the deobfuscation examples we consider have small teaching dimension.

---

[2]This view of a specification as an I/O oracle applies to many other contexts, including programming by demonstration and end-user programming, where the I/O oracle may be a human user.

- *Deductive Reasoning:* This engine is an SMT solver that performs two functions as noted above: (i) it generates candidate programs consistent with generated input-output examples; and (ii) it generates new inputs that distinguish between two non-equivalent programs in $\mathcal{C}_\mathcal{H}$ consistent with the generated input-output examples.

The above instance of sciduction, although motivated by the malware deobfuscation problem, can be applied in other program synthesis settings as well.

### 4.3 Theoretical Guarantees and Sample Results

The structure hypothesis is valid if the library of components defines a space of target programs $\mathcal{C}_\mathcal{H}$ containing one that is equivalent to the obfuscated program. If the structure hypothesis is valid, the sciductive approach sketched above and presented in [23] is *sound*. The program it generates is guaranteed to be the correct program (equivalent to the obfuscated program we start with). See Theorem 4 in [23] for details.

If, however, the structure hypothesis is invalid, then our approach could either report that the problem is unrealizable (i.e., there is no program synthesizable with the component library that matches the input-output examples) or it could output a program that is consistent with all seen input-output examples, but which is not the correct program. Figure 7 depicts the possible cases.

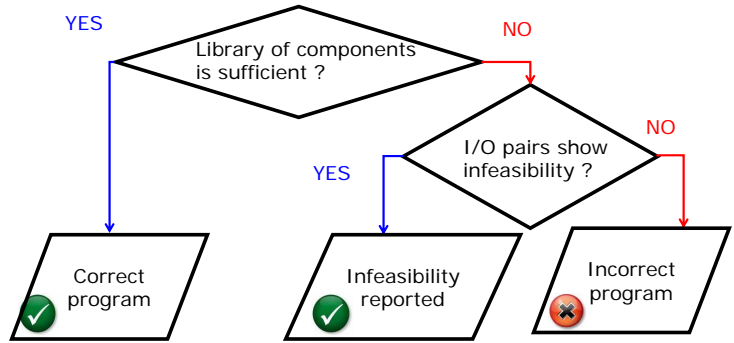

Figure 7: **Theoretical Guarantees for Program Synthesis Technique.**

In practice a range of heuristics are used to compute a sufficient component library *L*. Figure 8 shows two obfuscated programs and their deobfuscations computed using our approach. Both programs were deobfuscated in less than half a second.

## 5 Switching Logic Synthesis

Cyber-physical systems, which integrate physical dynamics and computational systems, are often conveniently modeled as multi-modal dynamical systems (MDSs). An MDS is a physical system (also known as a "plant") that can operate in different modes. The dynamics of the plant in each mode is known, and is usually

**P1:** Interchange the source and destination addresses.

**P2:** Multiply by 45

```
interchangeObs(IPaddress* src, IPadress* dest)
{ *src = *src ^ *dest;
  if (*src == *src ^ *dest)
    { *src = *src ^ *dest;
      if (*src == *src ^ *dest)
      { *dest = *src ^ *dest;
        if (*dest == *src ^ *dest)
        { *src = *dest ^ *src;
          return;
        }
        else
        { *src = *src ^ *dest;
          *dest = *src ^ *dest;
           return;
        }
      }
      else
        *src = *src ^ *dest;
    }
    *dest = *src ^ *dest;
    *src = *src ^ *dest;
    return;
}

interchange(IPaddress* src, IPadress* dest)
{
 *dest = *src ^ *dest;
 *src =  *src ^ *dest;
 *dest = *src ^ *dest;
 return;
}
```

```
int multiply45Obs(int y)
{ a=1; b=0; z=1; c=0;
   while(1) {
     if (a == 0) {
       if (b == 0) {
          y=z+y; a =~a; b=~b;c=~c;
          if (~c) break;
       }
       else {
          z=z+y; a=~a; b=~b; c=~c;
          if (~c) break;
       }
     }
     else {
       if (b == 0) { z=y<<2; a=~a; }
       else {
         z=y << 3;
         a=~a; b=~b;
       }
     }
   }
   return y;
}

multiply45(int y)
{
   z = y << 2;
   y = z + y;
   z = y << 3;
   y = z + y;
   return y;
}
```

Figure 8: **Some deobfuscation benchmarks presented in [23].** For both benchmarks (a) and (b), the original obfuscated program is shown at the top and the resynthesized program generated by our system at the bottom.

specified using a continuous-time model such as a system of ordinary differential equations (ODEs). However, to achieve safe and efficient operation, it is typically necessary to switch between the different operating modes using carefully constructed *switching logic*: guards on transitions between modes. The MDS along with its switching logic consitutes a *hybrid system*. Manually designing switching logic so as to ensure that the hybrid system satisfies its specification can be tricky and tedious.

While several techniques for switching logic synthesis have been proposed (see, e.g., [3, 62, 27, 35, 39, 15, 61]), it remains quite challenging to handle systems with a combination of rich discrete structure (in the form of multiple modes) and complex non-linear dynamics within modes. We discuss one such representative switching logic synthesis problem below.

## 5.1 The Problem

We consider the switching logic synthesis problem for *safety*. See [24] for formal definitions; an informal and more intuitive presentation is made here. A safety property can be viewed as a subset of evaluations to the $n$ continuous state variables (i.e., a subset of $\mathbb{R}^n$); each evaluation is a *safe state*. A hybrid system is safe from a set of initial states if every reachable state is a safe state.

The problem of interest is as follows:

> Given a safety property, a multimodal dynamical system (MDS), and a set of initial states, synthesize switching logic for the MDS so that the resulting hybrid system is safe.

We impose no constraints on the intra-mode continuous dynamics in the MDS, other than it be deterministic and locally Lipschitz at all points [24].

In this article, we model hybrid systems using the *hybrid automaton* formalism [1]. An example switching logic synthesis problem is the 3-gear automatic transmission system depicted in Figure 9 [30]. This example has seven modes. The transitions between modes are labeled with guard variables: $g_{ij}$ labels the transition from Mode $i$ to Mode $j$. Such a guard is termed an *entry guard* for Mode $j$ and an *exit guard* for Mode $i$.

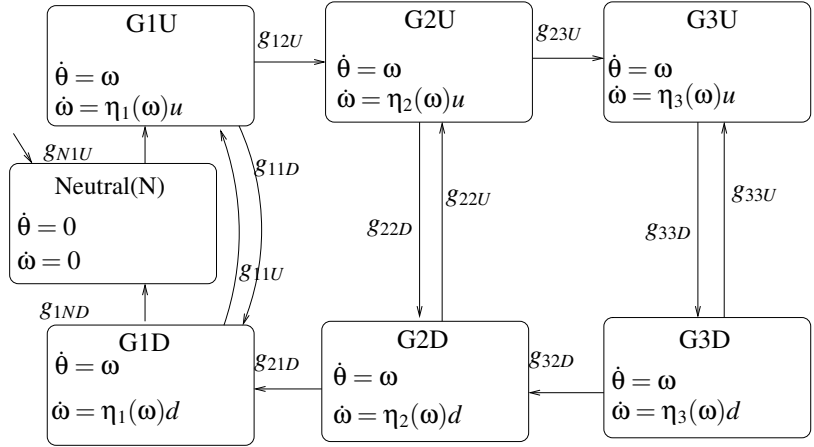

Figure 9: Automatic Transmission System

Note that for this example, the dynamics in each mode are *non-linear differential equations*. $u$ and $d$ denote the throttle in accelerating and deaccelerating mode. The transmission efficiency $\eta$ is $\eta_i$ when the system is in the $i$th gear, given by:
$$\eta_i = 0.99 e^{-(\omega - a_i)^2 / 64} + 0.01$$
where $a_1 = 10, a_2 = 20, a_3 = 30$ and $\omega$ is the speed. The distance covered is denoted by $\theta$. The acceleration in mode $i$ is given by the product of the throttle and transmission efficiency.

For simplicity, suppose that $u = 1$ and $d = -1$. Also, let the initial state be $\theta = 0, \omega = 0$. Suppose that the system must be designed to reach $\theta = \theta_{max} = 1700$ with $\omega = 0$. The synthesis problem is to find the guards between the modes such that the efficiency $\eta$ is high for speeds greater than some threshold, that is, $\omega \geq 5 \Rightarrow \eta \geq 0.5$. Also, $\omega$ must be less than an upper limit of 60. So, the safety property $\phi_S$ to be enforced would be
$$(\omega \geq 5 \Rightarrow \eta \geq 0.5) \wedge (0 \leq \omega \leq 60)$$
We will additionally assume that we know an initial overapproximation of the guards. Since the speed must reduce to 0 on reaching $\theta_{max}$, the guard $g_{1ND}$ is initialized to $\phi_S \wedge \theta = \theta_{max} \wedge \omega = 0$. All the other guards are initialized to $0 \leq \omega \leq 60$. Clearly every switching state must be a safe state.

Note that for the class of hybrid automata with nonlinear dynamics within modes, even reachability analysis is undecidable. Synthesizing safe switching logic is therefore undecidable too, unless additional assumptions are imposed. While we cannot expect to have a synthesis procedure that works in all cases, our experience is that it is possible to develop an approach that handles many cases arising in practice.

## 5.2 Our Approach

We adopt once again a sciductive approach to the controller synthesis problem, with the following elements:

- *Structure Hypothesis:* The essence of the structure hypothesis is to impose a particular syntactic form on the guards of the hybrid system: that the guards are hyperboxes. More precisely, the structure hypothesis includes the following two properties:

  1. The safe switching logic, if one exists, has all guards as *n*-dimensional hyperboxes with vertices lying on a known discrete grid.[3]
  2. For each mode, if all exit guards and all but one entry guard are fixed as hyperboxes, then for the remaining entry transition to that mode, the safe switching states constitute a hyperbox on the above-mentioned discrete grid.

  Since the set of initial states is also a particular kind of guard (on the "transition" that initializes the hybrid system), the structure hypothesis will also apply to the set of initial states.

  While the above structure hypothesis may not be valid for general multi-modal dynamical systems, it can be proved valid under two additional properties: (i) the continuous dynamics within a mode is such that state variables vary *monotonically* within a mode [24], and (ii) the discrete grid reflects the finite-precision with which values of continuous system variables can be recorded.

  To summarize, $\mathcal{C}_{\mathcal{H}}$ is the set of all hybrid automata in which the guards satisfy the above structure hypothesis.

- *Inductive Inference:* This routine is an algorithm to learn hyperboxes in $\mathbb{R}^n$ from labeled examples. An example is a point in $\mathbb{R}^n$. Its label is positive if the point is inside the box, and negative otherwise.

  More specific to our problem context, the learning problem is as follows. We are given a mode with its associated entry and exit guards. These guards are assumed to be overapproximate hyperboxes — the guards of a safe switching logic, if one exists, are subsets of the corresponding overapproximate guards. Given an entry guard, which could contain both safe and unsafe switching states, we want to infer a hyperbox that contains only the safe switching states and none of the unsafe switching states.

  If the structure hypothesis is valid, such an entry guard exists and our inductive inference routine can find it. The idea is to view safe switching states as positive examples and unsafe switching states as negative examples. The diagonally opposite corners of this hyperbox can then be found using binary search from the corners of the starting overapproximate hyperbox, assuming points in the hyperbox can be labeled as safe/unsafe (positive/negative). The search terminates when we have found the lower and upper diagonal corners as positive examples with their "immediate outer neighbours" as negative examples; for further details, see the hyperbox learning problem discussed by Goldman and Kearns [20].

  The positive/negative labels on states, required by the inductive routine, are generated by a deductive engine, as described below.

- *Deductive Reasoning:* In order to label a switching state *s* for a mode *m* as safe or unsafe, we need a procedure to answer the following question: if we enter *m* in state *s* and follow its dynamics, will the trajectory visit only safe states until some exit guard becomes true?

---

[3]Recall that a hyperbox corresponds to a conjunction of interval constraints over the continuous variables. The requirement for the vertices of the hyperbox to lie on a discrete grid is equivalent to requiring the constant terms in the hyperbox to be rational numbers with known finite precision.

This is a reachability analysis problem for purely continuous systems modeled as a system of ordinary differential equations (ODEs) with a single initial condition. This problem is known to be undecidable in general [48].

However, in practice, this reachability problem can be solved for many kinds of continuous dynamical systems (including the intra-mode dynamics for the example shown in Fig. 9) using state-of-the-art techniques for *numerical simulation* (see, e.g., [47]). Thus, the deductive engine in our sciductive approach is a numerical simulator that can handle the dynamics in each mode of the multi-modal dynamical system. The numerical simulator must be *ideal*, in that it must always return the correct YES/NO answer to the above reachability question.

The reader might wonder why a numerical simulator is termed as a deductive engine. Indeed, on the surface a numerical simulator seems quite different from a deductive theorem prover. However, on closer inspection one finds that both procedures employ similar deductive reasoning: they both solve systems of constraints using axioms about underlying theories or rules of inference, and they both involve the use of rewrite and simplification rules.

Our overall approach to switching logic synthesis for safety properties [24] operates within a fixpoint computation loop that initializes each guard with an overapproximate hyperbox, and then iteratively shrinks entry guards using the hyperbox learning algorithm that selects states, queries the simulator for labels, and then infers a smaller hyperbox from the resulting labeled states.

### 5.3 Theoretical Guarantees

If the structure hypothesis is valid and we have an ideal numerical simulator, our approach to switching logic synthesis for safety properties [24] is *sound and complete*. This follows from three aspects: (i) the initialization of each guard with an *overapproximate hyperbox*; (ii) the structure hypothesis that ensures that the safe switching states in each iteration will form a hyperbox, and (iii) the learning algorithm that yields a hyperbox for each guard at each iteration that excludes all negative examples (unsafe switching states) and includes all positive examples (safe switching states).

However, if the structure hypothesis is not valid, or if the numerical simulator is non-ideal, then our approach cannot be guaranteed to be sound or complete. For this reason, if one cannot prove or otherwise reasonably assume the structure hypothesis to hold for the class of systems of interest, and the simulator to be ideal, then one must separately formally verify that the synthesized system satisfies the safety property. The numerical simulator could also be replaced by an alternative reachability oracle.

### 5.4 Sample Result

We present here a sample result obtained for the automatic transmission example of Fig. 9. Our procedure combines learning hyperboxes (intervals, in this example) with a Matlab-based numerical simulator within an overall fixpoint computation loop. The outer loop starts with all guards set to the safety property $\phi_S$ (defined above) and iteratively shrinks each guard using the binary search-based inductive procedure described above. The final set of guards obtained after fixpoint computation are as follows.

$$g_{N1U}, g_{11U} : 0 \leq \omega \leq 16.70$$
$$g_{12U}, g_{22U} : 13.29 \leq \omega \leq 26.70$$
$$g_{23U}, g_{33U} : 23.29 \leq \omega \leq 36.70 \, , \, g_{33D} : 23.29 \leq \omega \leq 36.70$$
$$g_{32D}, g_{22D} : 13.29 \leq \omega \leq 26.70$$
$$g_{21D}, g_{11D} : 0 \leq \omega \leq 16.70 \, , \, ; \, g_{1ND} : \theta = \theta_{max} \wedge \omega = 0 \quad (3)$$

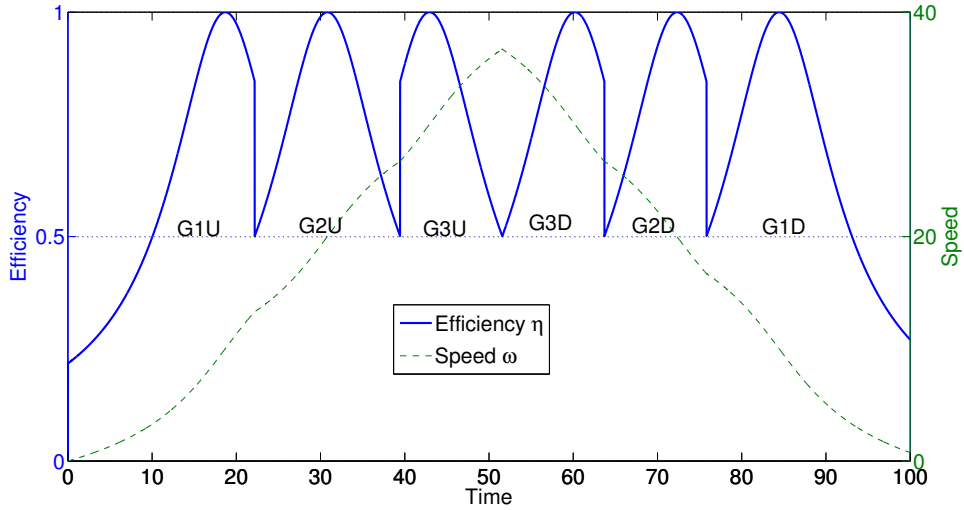

Figure 10: Transmission efficiency and speed with changing gears

Slightly modifying the safety property to require that the system spends at least 5 seconds in each of the six gear modes, we obtain the following modified set of guards:

$$g_{N1U} : \omega = 0 \,,\, g_{11U} : \omega = 0$$
$$g_{1ND} : \theta = \theta_{max} \wedge \omega = 0 \,,\, g_{12U} : 13.29 \leq \omega \leq 23.42$$
$$g_{11D} : 1.31 \leq \omega \leq 16.70 \,,\, g_{23U} : 26.70 \leq \omega \leq 33.42$$
$$g_{22D} : \omega = 26.70 \,,\, g_{33D} : \omega = 36.70$$
$$g_{32D} : 16.58 \leq \omega \leq 26.70 \,,\, g_{33U} : 23.29 \leq \omega \leq 33.42$$
$$g_{21D} : 1.31 \leq \omega \leq 16.70 \,,\, g_{22U} : 13.29 \leq \omega = 23.42 \quad (4)$$

The plot of the behavior of the transmission system when it is made to switch from Neutral mode through the six gear modes and back to the Neutral mode is shown in Figure 10. The efficiency $\eta$ is always greater than 0.5 when the speed is higher than 5 and we spend atleast 5 seconds in the six gear modes. Starting from $\theta = 0, \omega = 0$, the synthesized system reaches $\theta = \theta_{max}$ with $\omega = 0$.

# 6 Conclusions and Future Directions

This paper posits that sciduction, a tight integration of induction and deduction with structure hypotheses, is a promising approach to addressing challenging problems in formal verification and synthesis. We have demonstrated some initial results in this regard, summarized in Table 1.

We conclude with some thoughts on further work on the sciductive approach and its applications.

**Structure Hypothesis Testing/Verification.** Recall that the soundness guarantees of sciduction only hold when the structure hypothesis is valid. A limitation of the current demonstrations of sciduction is that we currently do not have a systematic and general approach for checking the validity of the structure hypothesis. For example, in the program synthesis application of Sec. 4, how can we be sure that the library of components is sufficient to synthesize the program? As noted in Fig. 7, if the structure hypothesis does not hold, it is possible to output an incorrect program. In this case, testing the structure hypothesis requires checking equivalence of the generated program against the specification, which may be expensive. More effective and generally-applicable methods for testing the structure hypothesis are required.

| Application | $\mathcal{H}$ | $\mathcal{I}$ | $\mathcal{D}$ |
|---|---|---|---|
| Timing analysis (Sec. 3) | $w + \pi$ model & constraints | Game-theoretic online learning | SMT solving for basis path generation |
| Program synthesis (Sec. 4) | Loop-free programs from component library | Learning from distinguishing inputs | SMT solving for input/program generation |
| Switching logic synthesis (Sec. 5) | Guards as hyperboxes | Hyperbox learning from labeled points | Numerical simulation as reachability oracle |

Table 1: **Three Demonstrated Applications of Sciduction.** For each application, we briefly describe the structure hypothesis $\mathcal{H}$, the inductive inference engine $\mathcal{I}$, and the deductive procedure $\mathcal{D}$.

**Integrating Induction and Deduction.** Sciduction offers ways to integrate inductive reasoning into deductive engines, and vice-versa. It is intruiging to consider if SAT and SMT solvers can benefit from a sciductive approach — for example, using inductive reasoning to guide the solver for specific families of SAT/SMT formulas. Similarly, how can one effectively use deductive engines as oracles in learning algorithms? Are there new concept learning problems that can be effectively solved using this approach?

**New Applications.** An interesting direction is to take problems that have classically been addressed by purely deductive methods and apply the sciductive approach to them. For example, consider the problem of synthesis from LTL specifications. One practical challenge for this problem is in writing complete and consistent specifications, of which the environment assumptions are a large part. In recent work, we have demonstrated that environment assumptions can be mined from traces and counter-strategies [29]. It would be interesting to see if the synthesis algorithms themselves can be made more scalable using sciduction.

Sciduction can be used in generating abstractions or inductive invariants for verification. For example, we have recently used a combination of induction on decision trees (see [38]) and SMT-based ("term-level") model checking using UCLID [10] to perform conditional term-level abstraction of hardware designs [7]. Much remains to be explored in this area.

Controller synthesis for hybrid systems also remains an important domain with several applications. We have obtained some initial results on synthesizing switching logic for optimality, rather than just safety [25].

Another direction is to generalize the ideas used for timing analysis to other quantitative properties of cyber-physical systems, and also for verification problems at the hardware-software interface ("hardware-software verification"). In both settings, generating environment models can be quite challenging, and, from our experience with timing analysis, it appears that sciduction can be effectively brought to bear on these problems.

# Acknowledgments

This article is a result of ideas synthesized and verified (!) over the last few years in collaboration with several students and colleagues. In particular, Susmit Jha is a major contributor to this work, especially to Sections 4 and 5, which are part of his Ph.D. thesis research. Other collaborators include Jonathan Kotker and Alexander Rakhlin (Section 3), and Sumit Gulwani and Ashish Tiwari (Sections 4 and 5). This paper has benefited from feedback on talks on this work given by the author at several venues in 2009-11, including Princeton

University, UC Berkeley, University of Texas at Austin, University of Pennsylvania, Carnegie Mellon University, Stanford University, IIT Bombay, FMCAD 2010, and the 2010 Gigascale Systems Research Center (GSRC) annual meeting. Discussions with Dan Holcomb, Wenchao Li, and Dorsa Sadigh are also gratefully acknowledged.

The research reported here has been supported in part by several sponsors including the National Science Foundation (CNS-0644436, CNS-0627734, and CNS-1035672), Semiconductor Research Corporation (SRC) contracts 1355.001 and 2045.001, an Alfred P. Sloan Research Fellowship, the Hellman Family Faculty Fund, the Toyota Motor Corporation under the CHESS center, and the Gigascale Systems Research Center (GSRC) and MultiScale Systems Center (MuSyC), two of six research centers funded under the Focus Center Research Program (FCRP), a Semiconductor Research Corporation entity.